# Evidence of Predicting Early Signs of Corporate Bankruptcy Using Financial Ratios in the Indian Landscape

Abhi Bansal*, Adit Chopra*, Aryaman Wadhwa*

Shaheed Sukhdev College of Business Studies, University of Delhi

**ABSTRACT**

*Corporate Bankruptcy impacts the functioning of the economy adversely as it impacts the shareholders, financial lenders, operational lenders and government, thus making it pertinent to be able to predict when would a firm's performance deteriorate to a level where it has a high potential of filing for bankruptcy. This paper aims to study the impact of financial ratios in classifying companies into the two categories; bankrupt and non-bankrupt using logistical regression model considering the impact at two levels; one year before bankruptcy and two years before the bankruptcy. The research takes a sample of 90 public listed Indian companies (45 each for bankrupt and non-bankrupt firms) where the bankrupt companies filed for bankruptcy with NCLT in 2018 and 2019. It showed that it does not need many ratios to be able to anticipate potential business bankruptcy. The bankruptcy probability model is constructed using Profitability, Leverage and Efficiency ratio variables. The study proves that the accuracies of the classification model are 81.4% and 85.1% respectively for one year and two years before the bankruptcy.*

## 1. INTRODUCTION

Corporate Bankruptcy models provide an estimation of the probability of a firm going bankrupt using a set of covariates such as financial ratios. This is often referred to as Business Failure Prediction (BFP). The investigations of bankruptcy prediction research (Altman 1983; Ballantin 1992; D'Aveni 1989; Dugan and Zavgren 1989; Koh and Killough 1990; Pech and Alistair 1993; Shumway 2001; Chava and Jarrow 2004; Bunyaminu and Mohammed 2012) usually classify the companies in a binary method into distressed or non-distressed companies.

It is in the interest of investment funds, borrowing organizations and governments across the world to predict corporate bankruptcy. BFP helps avoid lending to (or investing in) businesses which are likely to fail, gives an early identification of at-risk business to government regulatory bodies, and provides scoring models with higher accuracy to rating agencies.

For investors, firms engaging in distress motivated restructuring show signs of performance improvement (Jensen, 1989; Whitaker, 1999). For managers, the right types of preventative measures must be taken when they recognize that their firm is in distress. The restructuring strategies chosen should be appropriate for the stage of corporate lifecycle that the company is in currently and hopefully, turn the firm around from distress. For policymakers, there is a risk of managers of distress firms taking the wrong and lifecycle-inconsistent restructuring decisions that can be detrimental to macroeconomic and financial stability.

This research paper aims to create a BFP model using logistic regression as the primary research methodology which can help predict whether a firm is at risk of going bankrupt in the next two years.

## 2. LITERATURE REVIEW

Financial distress prediction is a topic of great eminence for researchers, institutions and individuals across the world because of its incredible significance to companies, the economy, and all other concerned parties *(Wanke et al. 2015)*.

We reviewed major contributions to the field of Business Failure Prediction (BFP) by researchers across the globe. Initially, we explored traditional distress prediction models. Nowadays, more advanced statistical and computational techniques such as artificial neural networks, survival analysis and other machine learning based methodologies are utilised to provide a more accurate prediction methodology. However, our approach deals with a much more preliminary and theoretical technique of logistic regression. Thus, before we deep dive into our research methodology, analysis and interpretation of results, we first review existing literature in the field of BFP with an emphasis on traditional distress prediction models.

### *Traditional Distress Prediction Models*

The existing literature on financial distress prediction is extensive yet fragmented when it comes to utilization of independent variables. *Beaver (1966)* used univariate discriminant analysis to compare the ratios of failed and non-failed firms, and numerous bankruptcy prediction models based on this have been developed and tested by researchers. In 1968, one of the more well-known and widely used works came into existence which aimed at extending the work of *Beaver: Altman.* Altman aimed at employing the use of multiple discriminant analysis (MDA) to identify certain explanatory independent variables that could predict financial distress with a greater accuracy. Since *Altman (1968)*, multiple researchers in this field have utilized discriminant analysis to predict financial distress and bankruptcy. Most notably, *Deakin (1972), Grice and Ingram (2001), and Agarwal and Taffler (2007)*.

Many researchers have extended the Altman Z-Score model at a more geographically local extent. *Rim and Roy (2014)* performed an empirical research by testing the original Altman Z-Score for Lebanon based manufacturing companies and results indicated the viability of this model to predict financial distress of the Lebanese manufacturing sector extensively.

Other studies by *Li and Rahgozar (2012)* and *Ilahi et al. (2015)* further re-enforce this assertion. This by no means indicates that there doesn't exist any criticism for the Altman Z-

Score model. *Almamy et al. (2016)* found that the prediction accuracy of the original Z-score model declined with time for the UK market, especially during the global financial crisis.

The MDA technique has its fair share of criticism from a theoretical standpoint too. This is primarily due to of its restrictive assumptions about multivariate normality and the independence of explanatory variables *(Ohlson 1980).*

Following up on this assertion, *Ohlson (1980)* proposed a new model based on logit analysis with a set of nine accounting ratios. This piece of literature opened up the field to a new methodology of analysing probability of default; Logistic Regression.

This resulted in an increase in the number of studies which used logit regression and an improvement in accuracy of new developed BFP models *(Campbell et al. 2008; Sun et al. 2014; Jones et al. 2015, 2017).* Later, *Zmijewski (1984)* used probit regression analysis and developed a three-variable financial distress prediction model, which was tested by many researchers, such as *Wu et al. (2010)* and *Kleinert (2014)*. *Shumway (2001)* developed an extension of financial distress prediction models which criticized static bankruptcy prediction techniques used previously and developed a discrete hazard model, adding market-based variables, which led to an increase in the overall classification accuracy of the model. His model was tested many researchers, such as *Campbell et al. (2008*) and *Bonfim (2009)*. Subsequently, researchers such as *Chava and Jarrow (2004)* and *Agarwal and Taffler (2008)* stated that variables based on market data that reflect internal as well as external information increase the overall predictability and hence accuracy of financial distress prediction models.

One of the core drawbacks of the above explored default prediction models is the lack of a strong theoretical framework. Let's explore this point. The study of *Altman (1968)* was developed with limited data available, and was focused mainly on searching for the right variable. In order to counter this problem, a D-Score model was proposed by *Blums (2003).* This model explored explanatory variables which has a strong accounting and market-based conceptual framework. In further studies, multiple new variables were added into financial distress prediction models to enhance the robustness of the model from a theoretical perspective.

*Tykvová and Borell (2012)* took into account liquidity, profitability and solvency ratio. *Korol (2013)* used profitability, liquidity and activity ratios with a strong theoretical backing for choosing the same. In recent years, in addition to these statistical-based techniques, researchers have been exploring more computationally advanced methodologies such as SVM, genetic algorithms, decision tress and ANNs.

Most of the BFP literature has relied on relatively simple statistical prediction methodologies, as they are efficient predictors of financial distress *(Jones et al. 2017)*; therefore, we also restricted our study to simple statistics-based techniques.

## 3. RESEARCH METHODOLOGY

## 3.1 Data Analysis Technique: Logistic Regression

### 3.1.1 Overview

In setting up the logistic regression model, we first establish the fundamental model for any multiple regression analysis. The dependent variable is assumed as a linear combination of a set of independent variables. If dependent variable is Y, and a set of "n" independent variables are $X_1, X_2,..., X_n$, the Logit model is:

$$Y = \beta_0 + \beta_1 X_1 + \beta_2 X_2 + ... + \beta_n X_n + \varepsilon = \beta_0 + \sum_{j=1}^{n} \beta_j X_j$$

Where $\beta_0$ is the expected value of Y when X's set 0.
$\beta_j$ is the regression coefficient for each corresponding predictor variable $X_j$.
$\varepsilon$ is the error of the prediction.

We define $\pi(x)$ as the probability that Y = 1. Similarly, $1-\pi(x)$ is the probability that Y = 0. These probabilities are written in the following form:

$$\pi(x) = P(Y = 1 \mid X_1, X_2,..., X_n)$$
$$1-\pi(x) = P(Y = 0 \mid X_1, X_2,..., X_n)$$

The model for the natural logarithm of $\frac{\pi(x)}{1-\pi(x)}$ is:

$$\ln \frac{P(Y=1 \mid X1, X2,...,Xn)}{1-P(Y=1 \mid X1, X2,...,Xn)} = \ln \frac{\pi(x)}{1-\pi(x)} = \beta_0 + \sum_{j=1}^{n} \beta_j X_j$$

The conditional mean is between 0 and 1.

Firstly, we must establish a technique for estimating the parameters. Least Squares is a method of parameter estimation in logistic regression. For a set of observations in the data $(x_i, y_i)$, the contribution to the least squares is $\pi(x_i)$, where $y_i = 1$, and $1-\pi(x_i)$, where $y_i = 0$. The following equation results for the contribution to the least squares for the observation $(x_i, y_i)$ is $\varsigma(x_i)$

$$\varsigma(xi) = \pi(xi)^{yi}[1 - \pi(xi)^{1-yi}]$$

The solution can be solving by using computer programs such as Stata and R. It performs the logistic regression analysis of the data for this study and will calculate the least square estimates.

## 3.2 MLE

Given samples $(xi, yi) \in R\ p \times \{0, 1\}$, i = 1, . . . n, we let p(xi) = P(yi = 1|xi), and assume:

$$\text{Log } p(x_i)/(1 - p(x_i)) = \beta^T x_i, i = 1,..,n$$

To construct an estimate β̂ of the coefficients, we will use the principle of maximum likelihood. I.e., assuming independence of the samples, the likelihood (conditional on xi, i = 1, . . . n) is

$$L(\beta) = \prod_{i:y_i=1} p(x_i) \cdot \prod_{i:y_i=0} (1 - p(x_i))$$
$$= \prod_{i=1}^{n} p(x_i)^{y_i} (1 - p(x_i))^{1-y_i}.$$

We will choose β̂ to maximize this likelihood criterion

Note that maximizing a function is the same as maximizing the log of a function (because log is monotone increasing). Therefore, β̂ is equivalently chosen to maximize the log likelihood

$$\ell(\beta) = \sum_{i=1}^{n} y_i \log p(x_i) + (1 - y_i) \log(1 - p(x_i)).$$

It helps to re-arrange this as,

$$\ell(\beta) = \sum_{i=1}^{n} y_i \big[\log p(x_i) - \log(1 - p(x_i))\big] + \log(1 - p(x_i))$$
$$= \sum_{i=1}^{n} y_i \log\left(\frac{p(x_i)}{1 - p(x_i)}\right) + \log(1 - p(x_i)).$$

Finally, plugging in for log(p(xi)/(1 − p(xi))) = x T i β and using 1 − p(xi) = 1/(1 + exp(x T i β)), i = 1, .

$$\ell(\beta) = \sum_{i=1}^{n} y_i (x_i^T \beta) - \log(1 + \exp(x_i^T \beta)).$$

You can see that, unlike the least squares criterion for regression, this criterion (β) does not have a closed-form expression for its maximizer (e.g., try taking its partial derivatives and setting them equal to zero). Hence, we have to run an optimization algorithm to find β0

Somewhat remarkably, we can maximize this by running repeated weighted least squares regressions! This is actually just an instantiation of Newton's method. Applied to the criterion, we refer to it as iteratively reweighted least squares or IRLS

**3.3 Decision Boundary**

Suppose that we have formed the estimate β̂ of the logistic coefficients, as discussed in the last section. To predict the outcome of a new input x ∈ R p, we form

$$p(x) = \frac{\exp(\beta^T x)}{1 + \exp(\beta^T x)}$$

and then predict the associated class according

$$f(x) = \begin{cases} 0 & p(x) \leq 0.5 \\ 1 & p(x) > 0.5 \end{cases}$$

Equivalently, we can study the log odds and predict the associated class

$$logit\ p(x) = \beta^T x$$

The set of all x ∈ R p such that

$$\beta^T x = \beta_1^T x_1 + \cdots + \beta_p^T x_p = 0$$

Is called the decision boundary between classes 0 and 1. On either side of this boundary, we would predict one class or the other

Remembering the intercept, we would rewrite the decision boundary as

$$\beta_0 + \beta_1^T x_1 + \cdots + \beta_p^T x_p = 0$$

This is a point when p = 1, it is a line when p = 2, and in general it is a (p − 1)-dimensional subspace. We would therefore say that logistic regression has a linear decision boundary; this is because the above equation is linear in *x*

### 3.4 Inference

A lot of the standard machinery for inference in linear regression carries over to logistic regression. We can solve for the logistic regression coefficients $\hat{\beta}$ by performing repeated weighted linear regressions; hence we can simply think of the logistic regression estimates $\hat{\beta}$ as the result of a single weighted linear regression—the last one in this sequence (upon convergence). Confidence intervals for $\beta_j$, j = 1, . . . p, and so forth, are then all obtained from this weighted linear regression perspective.

### 3.5 Data Collection Methodology

The experimental dataset has been divided into a train set and a test set. The model has been trained on the former and its accuracy has been measured on the latter. The dataset consists of both bankrupt and non-bankrupt firms so that the model is robust and extensive. The dataset consists of 45 firms which filed for bankruptcy under the NCLT in FY19 (Bankrupt) and 45 similar firms listed on the Nifty 100 (Non-Bankrupt). This dataset of 45 bankrupt and non-bankrupt firms each has been divided in the ratio 70:30 into the train and test set respectively. All the requisite data points for the 90 firms have been extracted from Capitaline Plus database.

## 3.6 Independent Variable Selection

Table 1: Variables Selected

| S. No. | Category | Variable |
|---|---|---|
| 1. | Profitability Ratios | EBIT Margin |
| 2. | | Return on Equity |
| 3. | | Return on Assets |
| 4. | Liquidity Ratios | Current Ratio |
| 5. | Leverage Ratios | Debt Ratio |
| 6. | | Debt to Equity Ratio |
| 7. | Altman Ratios | A = Working Capital / Total Assets |
| 8. | | B = Retained Earnings / Total Assets |
| 9. | | C = Earnings Before Interest and Tax / Total Assets |
| 10. | | D = Market Value of Equity / Total Liabilities |
| 11. | | E = Sales / Total Assets |

Karels and Prakash suggested a careful selection of ratios to be used in the development of bankruptcy prediction model. A set of covariates used in this study includes a combination of financial ratios and market variables. In financial reporting analysis, suggest five factors for evaluation enterprise financial failure. Financial ratios have been widely used in explaining the possibility of business financial distress. The above table enlists the ratios we have considered in our analysis. Building up on Karels and Prakash's experiments, we have added certain Altman ratios and combined certain valuation and efficiency ratio to better capture the causes indicting probability of bankruptcy/default for corporates.

## 3.7 Reducing the number of financial ratios

We have considered the p-value of each financial ratio and have included variables that are significant at 5% level of significance.

Table 2: Reducing the number of financial ratios using p-value

| Variable | Coefficient | Std. Err. | z | P>|z| | [0.025 | 0.975] |
|---|---|---|---|---|---|---|
| EBIT Margin | -0.1126 | 1.5516 | -0.0726 | 0.9421 | -3.1538 | 2.9285 |
| RoE | -1.6191 | 1.5753 | -1.0278 | 0.304 | -4.7067 | 1.4684 |
| RoA | -145.683 | 62.3131 | -2.3379 | 0.0194 | -267.814 | -23.5512 |
| Current Ratio | -3.8754 | 1.8796 | -2.0618 | 0.0392 | -7.5594 | -0.1914 |
| D/E Ratio | 0.0904 | 0.2022 | 0.447 | 0.6548 | -0.3059 | 0.4866 |
| Debtors Ratio | -0.067 | 0.0466 | -1.437 | 0.1507 | -0.1584 | 0.0244 |
| Working Capital/Total Assets | 6.7881 | 3.7049 | 1.8322 | 0.0669 | -0.4733 | 14.0495 |
| EBIT/Total Assets | 56.3248 | 40.8666 | 1.7383 | 0.1681 | -23.7723 | 136.4219 |
| Sales/Total Assets | 4.0873 | 2.2615 | 1.8074 | 0.0707 | -0.3451 | 8.5197 |

The final variables used for building up the logistic regression model are; RoA, Current Ratio, Working Capital/Total Assets and Sales/Total Assets.

## 4. EMPIRICAL RESEARCH

In this section, the study first performs descriptive statistic of the sampling and Covariates, and follows by the construction of business failure prediction model based on Logistic Regression Model and then presents the analysis of empirical results.

Table 3: Descriptive Statistics for 2017

| Variables | Bankruptcy | RoA | Current Ratio | Working Capital/Total Assets | Sales/Total Assets |
|---|---|---|---|---|---|
| Mean | 0.5 | -0.00937 | 1.2808889 | 0.1526246 | 0.699243 |
| Standard Error | 0.053 | 0.01769 | 0.1161198 | 0.0352389 | 0.075464 |
| Median | 0.5 | 0.017015 | 1.01 | 0.1249259 | 0.47029 |
| Standard Deviation | 0.5028 | 0.167826 | 1.1016093 | 0.3343053 | 0.71591 |
| Kurtosis | 0.25281 | 0.028166 | 1.213543 | 0.1117601 | 0.512527 |
| Skewness | -2.046 | 11.0168 | 18.600427 | 14.526955 | 14.94245 |
| Range | 1.8E-17 | -2.427978 | 3.5550543 | -2.054057 | 3.017299 |
| Minimum | 1 | 1.22855 | 8.09 | 2.7752574 | 5.060267 |
| Maximum | 0 | -0.932435 | 0.22 | -1.875253 | 0.022706 |

Table 4: Descriptive Statistics for 2018

| Variables | Bankruptcy | RoA | Current Ratio | Working Capital/Total Assets | Sales/Total Assets |
|---|---|---|---|---|---|
| Mean | 0.5 | -0.1241 | 1.314667 | 0.1062332 | 0.62345432 |
| Standard Error | 0.053 | 0.05992 | 0.136122 | 0.0417745 | 0.06196602 |
| Median | 0.5 | 0.01101 | 0.945 | 0.0743867 | 0.45158506 |
| Standard Deviation | 0.5028 | 0.56844 | 1.291366 | 0.3963079 | 0.58786128 |
| Kurtosis | 0.25281 | 0.32313 | 1.667625 | 0.15706 | 0.34558089 |
| Skewness | -2.04598 | 24.2106 | 9.444862 | 23.932059 | 4.37316582 |
| Range | 1.8E-17 | -4.4547 | 2.799505 | -3.297989 | 1.79752684 |
| Minimum | 1 | 4.65492 | 7.51 | 3.5596329 | 3.27132868 |
| Maximum | 0 | -3.6566 | 0.14 | -2.593105 | -0.0038754 |

In order to obtain unbiased and efficient parameters, the independent variables should not be multicollinear i.e. the there should be no perfect correlation between any independent variables. We perform the correlation test to test for multicollinearity. The results are as follows;

Table 5: Correlation Matrix for 2017

| | Bankruptcy | RoA | Current Ratio | Working Capital/ Total Assets | Sales/Total Assets |
|---|---|---|---|---|---|
| Bankruptcy | 1.000 | | | | |
| RoA | -0.604 | 1.000 | | | |
| Current Ratio | -0.304 | 0.234 | 1.000 | | |
| Working Capital/Total Assets | 0.118 | 0.086 | 0.293 | 1.000 | |
| Sales/Total Assets | -0.067 | 0.237 | -0.100 | -0.106 | 1.000 |

Table 6: Correlation Matrix for 2018

|  | Bankruptcy | RoA | Current Ratio | Working Capital/ Total Assets | Sales/Total Assets |
|---|---|---|---|---|---|
| **Bankruptcy** | 1.000 |  |  |  |  |
| **RoA** | -0.379 | 1.000 |  |  |  |
| **Current Ratio** | -0.297 | 0.123 | 1.000 |  |  |
| **Working Capital/Total Assets** | -0.066 | 0.191 | 0.489 | 1.000 |  |
| **Sales/Total Assets** | -0.205 | 0.205 | -0.169 | -0.159 | 1.000 |

A logistic model was fitted to the data to test the research hypothesis regarding the relationship between the likelihood that a company is at risk for going bankrupt in the next two years and the Return on Assets, Current Ratio, ratio between Working Capital and Total Assets, and ratio between Sales and Total Assets. The logistic regression analysis was carried out by the logistic procedure in Python 3.8.

According to the model, for two years prior to the year of predicting bankruptcy, the log of the odds of a company being at risk for going bankrupt in the next two years was negatively related to return on assets (p = 0.0012) and current ratio (p = 0.0243) and positively related to ratio between Working Capital and Total Assets (p = 0.0176) and ratio between Sales and Total Assets (p = 0.0103). In other words, the higher the return on assets and current ratio, the less likely it is that a company will go bankrupt in the next two years; and the higher the ratio between working capital and total assets, and ratio between sales and total assets, the more likely it is that a company will go bankrupt in the next two years.

For one year prior to the year of predicting bankruptcy, the log of the odds of a company being at risk for going bankrupt in the next two years was negatively related to return on assets (p = 0.0000) and current ratio (p = 0.0114) and positively related to ratio between Working Capital and Total Assets (p = 0.0040). In other words, the higher the return on assets and current ratio, the less likely it is that a company will go bankrupt in the next year; and the higher the ratio between working capital and total assets, and ratio between sales and total assets, the more likely it is that a company will go bankrupt in the next year.

For the present data, these relationships are demonstrated in table given below:

Table 7: Output of Logistic Regression Model for 2017

|  | Coefficient | Std. Err. | z | P>\|z\| | [0.025 | 0.975] |
|---|---|---|---|---|---|---|
| **RoA** | -81.0931 | 25.0742 | -3.2341 | (0.0012)** | -130.2376 | -31.9486 |
| **Current Ratio** | -2.1581 | 0.9578 | -2.2531 | (0.0243)* | -4.0354 | -0.2808 |
| **Working Capital/Total Assets** | 6.9681 | 2.9354 | 2.3738 | (0.0176)* | 1.2148 | 12.7214 |
| **Sales/Total Assets** | 3.1037 | 1.2100 | 2.5651 | (0.0103)* | 0.7322 | 5.4752 |

Graph 1: ROC Curve for Logistic Regression Output for 2017

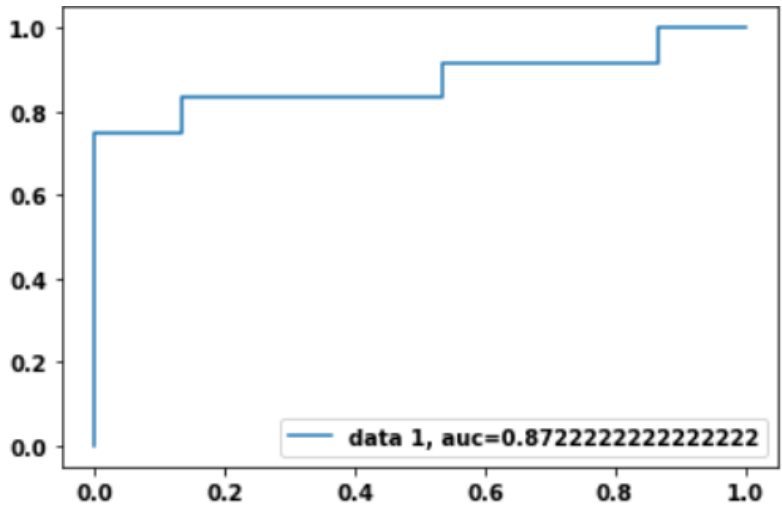

Graph 2: Confusion Matrix for Logistic Regression Output for 2017

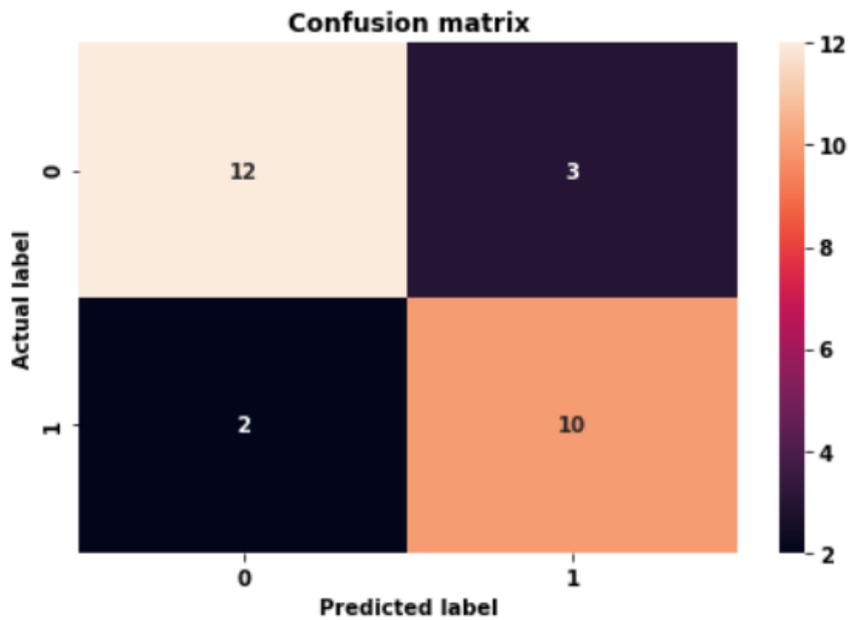

Table 8: Accuracy Table for 2017

|  | Bankrupt | Non-Bankrupt | Total |
| --- | --- | --- | --- |
| **Bankrupt** | 12 | 3 | 15 |
| **Non-Bankrupt** | 2 | 10 | 12 |
| **Total** | 14 | 13 | 27 |

Table 9: Robustness Check for Logistic Regression Output for 2017

| Measure | Value |
| --- | --- |
| Accuracy | 0.814 |
| Precision | 0.769 |
| Recall | 0.833 |

Table 10: Output of Logistic Regression Model for 2018

|  | Coefficient | Std. Err. | z | P>|z| | [0.025 | 0.975] |
|---|---|---|---|---|---|---|
| **RoA** | -14.5803 | 3.4136 | -4.2713 | (0.0000)*** | -21.2707 | -7.8898 |
| **Current Ratio** | -0.8462 | 0.3346 | -2.5920 | (0.0114)* | -1.5020 | -0.1904 |
| **Working Capital/Total Assets** | 2.3726 | 0.8246 | 2.8773 | (0.0040)* | 0.7564 | 3.9888 |
| **Sales/Total Assets** | 0.5384 | 0.4220 | 1.2759 | (0.2020) | -0.2886 | 1.3654 |

Graph 3: ROC Curve for Logistic Regression Output for 2018

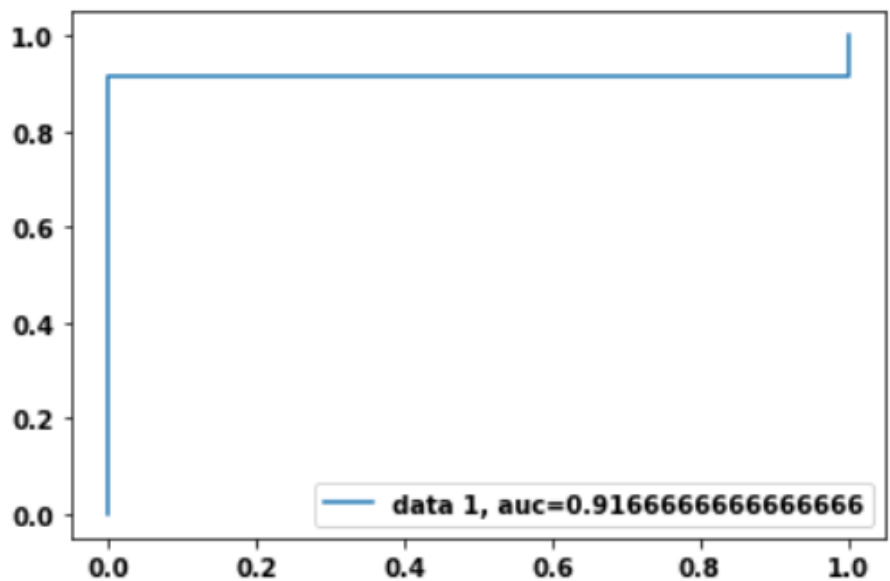

Graph 4: Confusion Matrix for Logistic Regression Output for 2018

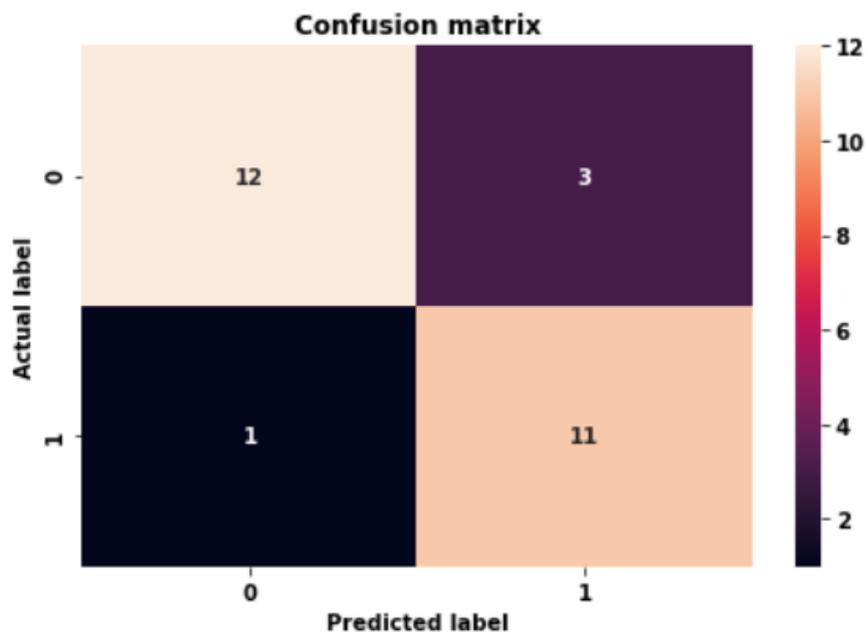

Table 11: Accuracy Table for 2018

|  | Bankrupt | Non-Bankrupt | Total |
|---|---|---|---|
| **Bankrupt** | 12 | 3 | 15 |
| **Non-Bankrupt** | 1 | 11 | 12 |
| **Total** | 13 | 14 | 27 |

Table 12: Robustness Check for Logistic Regression Output for 2018

| Measure | Value |
|---|---|
| Accuracy | 0.851 |
| Precision | 0.7857 |
| Recall | 0.916 |

According to our priori, the accuracy of the business failure prediction model increases as we approach closer to the date of bankruptcy. The accuracy of the logistic regression prediction model increase from 81.48% to 85.15% as we move from a time period of two years prior to bankruptcy (2017) to one year prior to bankruptcy (2018).

## 5. CONCLUSION

In this paper, the Indian companies that filed for bankruptcy in 2019 are employed as distressed data set. The sample was divided into train set samples and test set samples. This paper selected 45 distressed companies and 45 non-distressed companies for estimating samples data in 2019. The financial distress probability model is constructed using Return on Assets, Current Ratio, ratio between Working Capital and Total Assets, and ratio between Sales and Total Assets. We consider the robustness of model in prediction accuracy and in this study the accuracy of classification of the mode in overall accuracy of classification is 81.48% for prediction two years prior to the desired year of study and 85.15% for one year prior to the desired year of study.

## REFERENCES

Alaminos, D., del Castillo, A., & FernÂndez, M. Â. (2016). A Global Model for Bankruptcy Prediction. Department of Finance and Accounting, Universidad de MaÂlaga, MaÂlaga, Spain.

Altman, E.I., (1983), Corporate Financial Distress: A Complete Guide to Predicting, Avoiding and Dealing with Bankruptcy, Toronto: Wiley & Sons.

Altman, E. I., (1968). Financial Ratios, Discriminant Analysis and the Prediction of Corporate Bankruptcy. Journal of Finance.

Ballantine, J.W., Cleveland, F.W., & Koeller, C.T. (1992). Characterizing profitable and unprofitable strategies in small and large businesses. Journal of Small Business Management.

Bahnson, P.R & Bartley, J.W. (1992), The Sensitivity of failure prediction models to alternative definitions of failure, Advances in Accounting, 10, 255-278.

Beaver, W., (1967), "Financial Ratios as predictors of Failure" in Journal of Accounting Research.

Beaver, W., (1966). Financial Ratios as Predictors of Bankruptcy. Journal of Accounting Research.

Bunyaminu, A., & Bashiru, S. (2014). Corporate Failure Prediction: A Fresh Technique for Dealing Effectively With Normality Based On Quantitative and Qualitative Approach. International Journal of Financial Economics, 1-12.

Bunyaminu, A., & Issah, M. (2012). Predicting Corporate Failure of UK's Listed Companies: Comparing Multiple Discriminant Analysis and Logistic Regression. International Research Journal of Finance and Economics ISSN 1450-2887, 94(2012).

Chava, S. and R. A. Jarrow (2004). Bankruptcy Prediction with Industry Effects. Review of Finance.

Dakovic R., Claudia C., and Daniel B. (2007), Bankruptcy Prediction in Norway: A Comparison Study. Available online: http://www-m4.ma.tum.de/Papers/Czado/BankPred.pdf.

D'Aveni, R. (1989). The aftermath of organizational decline: A longitudinal study of the strategic and managerial characteristics of declining firms. Academy of Management Journal.

Dugan, M., & Zavgren, C. (1989). How a bankruptcy model could be incorporated as an analytical procedure. CPA Journal.

Fitzpatrick, P.J., (1932), A comparison of ratios of successful industrial enterprises with those of failed firms, Certified Public Accountant.

Fletcher, D & Goss, E. (1993), Forecasting with neural networks: An application using bankruptcy data, Information and Management, 24, 159-167.

Gilbert, L, K. Menon, and K, Schwartz (1990). Predicting bankruptcy for firms in financial distress. Journal of Business Finance and Accounting, Spring.

Lee, M.-C. (2014). Business Bankruptcy Prediction Based on Survival Analysis Approach. International Journal of Computer Science & Information Technology (IJCSIT), 103-119.

M. Ma'aji, M. e. (2018). Predicting Financial Distress among SMEs in Malaysia. European Scientific Journal, 91-102.

Ohlson, J. A. (1980). Financial Ratios and the Probabilistic Prediction of Bankruptcy. Journal of Accounting Research, 109-131.

Rao, N. V. (2013). ANALYSIS OF BANKRUPTCY PREDICTION MODELS AND THEIR EFFECTIVENESS: AN INDIAN PERSPECTIVE. Great Lakes Herald, 3-17.

Yang et. al (2018). PREDICTIONS, DISENTANGLING AND ASSESSING UNCERTAINTIES IN MULTIPERIOD CORPORATE DEFAULT RISK. Annuals of Applied Statistics.